\documentclass[twoside,twocolumn,9pt]{article}
\usepackage{bm}
\usepackage{multirow}
\usepackage{mathtools}
\usepackage{graphicx}
\usepackage{epstopdf}
\usepackage{wrapfig}
\usepackage{hyperref}
\usepackage{array} 
\usepackage{listings}
\usepackage[para,online,flushleft]{threeparttablex}
\usepackage{booktabs,dcolumn}
\usepackage{color}
\usepackage{physics}
\usepackage{amsmath}    
\usepackage{amssymb}
\usepackage{dcolumn}
\usepackage{extsizes}
\usepackage[super,sort&compress,comma]{natbib} 
\usepackage[version=3]{mhchem}
\usepackage[left=1.5cm, right=1.5cm, top=1.785cm, bottom=2.0cm]{geometry}
\usepackage{balance}
\usepackage{mathptmx}
\usepackage{sectsty}
\usepackage{graphicx} 
\usepackage{lastpage}
\usepackage[format=plain,justification=justified,singlelinecheck=false,font={stretch=1.125,small,sf},labelfont=bf,labelsep=space]{caption}
\usepackage{float}
\usepackage{fancyhdr}
\usepackage{fnpos}
\usepackage[english]{babel}
\addto{\captionsenglish}{%
  
}
\usepackage{array}
\usepackage{droidsans}
\usepackage{charter}
\usepackage[T1]{fontenc}
\usepackage[usenames,dvipsnames]{xcolor}
\usepackage{setspace}
\usepackage[compact]{titlesec}
\usepackage{hyperref}
\usepackage{cleveref}
\Crefname{figure}{Fig.}{Figs.}

\usepackage{epstopdf}

\definecolor{cream}{RGB}{222,217,201}

\begin{document}

\pagestyle{fancy}
\thispagestyle{plain}
\fancypagestyle{plain}{
\renewcommand{\headrulewidth}{0pt}
}

\makeFNbottom
\makeatletter
\renewcommand\LARGE{\@setfontsize\LARGE{15pt}{17}}
\renewcommand\Large{\@setfontsize\Large{12pt}{14}}
\renewcommand\large{\@setfontsize\large{10pt}{12}}
\renewcommand\footnotesize{\@setfontsize\footnotesize{7pt}{10}}
\renewcommand\scriptsize{\@setfontsize\scriptsize{7pt}{7}}
\makeatother

\renewcommand{\thefootnote}{\fnsymbol{footnote}}
\renewcommand\footnoterule{\vspace*{1pt}%
\color{cream}\hrule width 3.5in height 0.4pt \color{black} \vspace*{5pt}} 
\setcounter{secnumdepth}{5}

\makeatletter 
\renewcommand\@biblabel[1]{#1}            
\renewcommand\@makefntext[1]%
{\noindent\makebox[0pt][r]{\@thefnmark\,}#1}
\makeatother 
\renewcommand{\figurename}{\small{Fig.}~}
\sectionfont{\sffamily\Large}
\subsectionfont{\normalsize}
\subsubsectionfont{\bf}
\setstretch{1.125} 
\setlength{\skip\footins}{0.8cm}
\setlength{\footnotesep}{0.25cm}
\setlength{\jot}{10pt}
\titlespacing*{\section}{0pt}{4pt}{4pt}
\titlespacing*{\subsection}{0pt}{15pt}{1pt}

\fancyfoot{}
\fancyfoot[LO,RE]{\vspace{-7.1pt}\includegraphics[height=9pt]{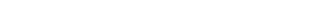}}
\fancyfoot[CO]{\vspace{-7.1pt}\hspace{13.2cm}\includegraphics{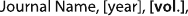}}
\fancyfoot[CE]{\vspace{-7.2pt}\hspace{-14.2cm}\includegraphics{head_foot/RF}}
\fancyfoot[RO]{\footnotesize{\sffamily{1--\pageref{LastPage} ~\textbar  \hspace{2pt}\thepage}}}
\fancyfoot[LE]{\footnotesize{\sffamily{\thepage~\textbar\hspace{3.45cm} 1--\pageref{LastPage}}}}
\fancyhead{}
\renewcommand{\headrulewidth}{0pt} 
\renewcommand{\footrulewidth}{0pt}
\setlength{\arrayrulewidth}{1pt}
\setlength{\columnsep}{6.5mm}
\setlength\bibsep{1pt}

\makeatletter 
\newlength{\figrulesep} 
\setlength{\figrulesep}{0.5\textfloatsep} 

\newcommand{\topfigrule}{\vspace*{-1pt}%
\noindent{\color{cream}\rule[-\figrulesep]{\columnwidth}{1.5pt}} }

\newcommand{\botfigrule}{\vspace*{-2pt}%
\noindent{\color{cream}\rule[\figrulesep]{\columnwidth}{1.5pt}} }

\newcommand{\dblfigrule}{\vspace*{-1pt}%
\noindent{\color{cream}\rule[-\figrulesep]{\textwidth}{1.5pt}} }

\makeatother

\twocolumn[
  \begin{@twocolumnfalse}
{\includegraphics[height=30pt]{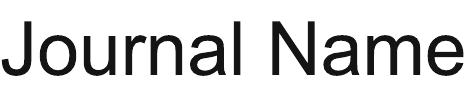}\hfill\raisebox{0pt}[0pt][0pt]{\includegraphics[height=55pt]{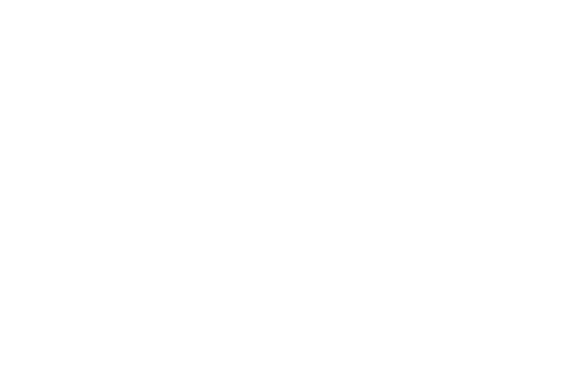}}\\[1ex]
\includegraphics[width=18.5cm]{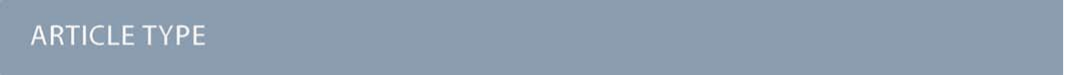}}\par
\vspace{1em}
\sffamily
\begin{tabular}{m{4.5cm} p{13.5cm} }

\includegraphics{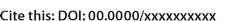} & \noindent\LARGE{\textbf{Large Vibrationally Induced Parity Violation Effects in CHDBrI$^+$ -- A Promising Candidate for Future Experiments$^\dag$}} \\
 & \vspace{0.3cm} \\

 & \noindent\large{Eduardus,\textit{$^{a}$} Yuval Shagam,\textit{$^{b}$} Arie Landau,\textit{$^{b}$} Shirin Faraji,\textit{$^{c}$} Peter Schwerdtfeger,\textit{$^{d}$} Anastasia Borschevsky,\textit{$^{a}$} and Luk\'a\v{s} F. Pa\v{s}teka$^{\ast}$\textit{$^{a,e}$}} \\

\includegraphics{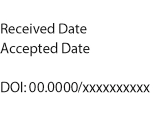} & \\

\end{tabular}

 \end{@twocolumnfalse} \vspace{0.6cm}

  ]

\renewcommand*\rmdefault{bch}\normalfont\upshape
\rmfamily
\section*{}
\vspace{-1cm}


\footnotetext{\textit{$^{a}$~Van Swinderen Institute for Particle Physics and Gravity (VSI), University of Groningen, Groningen, The Netherlands. E-mail: l.f.pasteka@rug.nl}}
\footnotetext{\textit{$^{b}$~Schulich Faculty of Chemistry, Solid State Institute and The Helen Diller Quantum Center, Technion-Israel Institute of Technology, Haifa, Israel }}
\footnotetext{\textit{$^{c}$~Zernike Institute for Advanced Materials, University
of Groningen, Groningen, The Netherlands }}
\footnotetext{\textit{$^{d}$~Centre for Theoretical Chemistry and Physics, The New Zealand Institute for Advanced Study, Massey University, Auckland, New Zealand }}
\footnotetext{\textit{$^{e}$~Department of Physical and Theoretical Chemistry, Faculty of Natural Sciences, Comenius University, Bratislava, Slovakia }}

\footnotetext{\dag~Electronic Supplementary Information (ESI) available. See DOI: 00.0000/00000000.}



\sffamily{\textbf{The isotopically chiral molecular ion CHDBrI$^+$ is identified as an exceptionally promising candidate for the detection of parity violation in vibrational transitions. 
The largest predicted parity-violating frequency shift reaches 1.8~Hz for the hydrogen wagging mode which has  a sub-Hz natural line width and its vibrational frequency auspiciously lies in the available laser range.
In stark contrast to this result, the parent neutral molecule is two orders of magnitude less sensitive to parity violation. The origin of this effect is analyzed and explained.
Precision vibrational spectroscopy of CHDBrI$^+$ is feasible as it is amenable to preparation at internally low temperatures and resistant to predissociation, promoting long interrogation times.\cite{landau} The intersection of these properties in this molecular ion places the first observation of parity violation in chiral molecules within reach. }}\\


\rmfamily 





Ever since Louis Pasteur stated in 1874 that life is dominated by dissymetric actions in our universe,\cite{Pasteur} the search for violation of fundamental symmetries in nature has captured physicists and chemists alike.\cite{Ginges2004,Jungmann2014} One of the greatest remaining puzzles in the life sciences concerns the origin of single handedness in nature, i.e. a distinct symmetry breaking that led to the dominance of left-handed amino acids and right-handed sugars \cite{Mason1988,Bonner1995,Blackmond2020} seen as a necessary condition for the existence of life on our planet. 

In 1975 Lethokov pointed out that the fundamental weak forces in physics, discovered by Wu in 1957\cite{Wu1} in an asymmetric $\beta$-decay experiment, lead to parity violation (PV) and thus to the breakdown of the left-right symmetry in chiral molecules.\cite{Letokhov75} While in typical biomolecules, the predicted PV effects in the electronic structure lead to a rather small energy difference of $\Delta E_\text{PV}<10^{-12}$ kJ/mol between two enantiomers, it is not yet clear if such tiny effects are responsible for biomolecular homochirality.\cite{Laerdahl2000a,Wesendrup2003} Naturally, there are many hypotheses on the origin of single handedness of life and the reader is referred to Refs. \citenum{Cintas2022} and \citenum{Crassous2022} for general reviews. Despite an ongoing debate on this issue, for the last several decades scientists have tried hard  to find experimental evidence for PV in chiral molecules\cite{CraChaSau05,DarStoZri10,Quack2022} -- so far with no success,\cite{DauMarAmy99,chardonnet:newexp} despite some questionable positive claims in the past.\cite{szabo1999,Wang2000,Wang2003}

\begin{figure}
\begin{center}
\includegraphics[width=8.6cm]{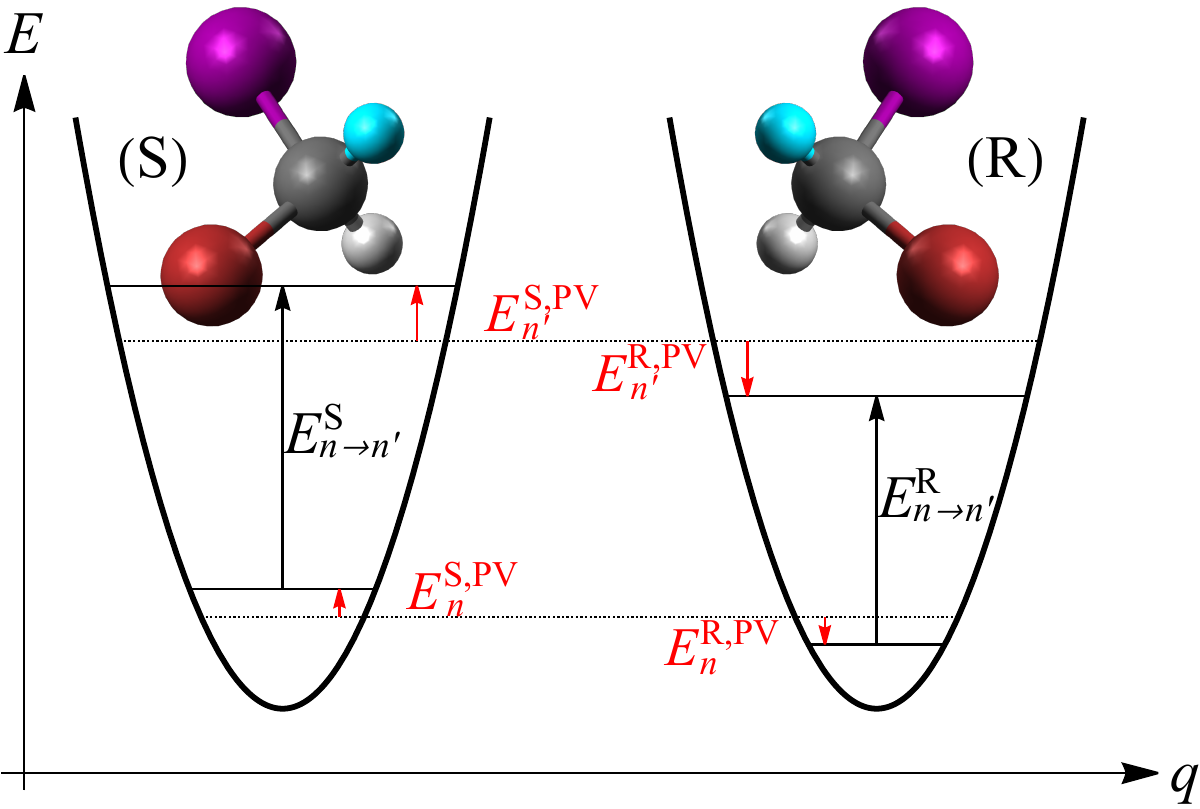}
\caption{Left (S) and right handed (R) enantiomers of CHDBrI$^+$ and response of their respective vibrational transitions to parity violation effects. Hydrogen is shown in light blue color.}
\label{fig:ChDBrI}
\end{center}
\end{figure}

The search for ideal candidates for PV measurements has focused on chiral molecules containing heavy atoms,\cite{BasSch03,BasSch04,FigSauSch10,Wormit2014,marit22} as PV contributions to the total energies scale approximately as $Z^5$ ($Z$ being the nuclear charge of the atom involved).\cite{Bouchiat:1974:physlettb,LaeSch99}
For vibrational PV measurements it was expected that large PV energy differences between the two enantiomers will translate into large PV effects in vibrational transitions. 
Molecules like CHDBrI, which by some chirality measures\cite{Mislow1992} have (almost) no chirality, and, furthermore, do not contain heavy metals, have been investigated in two earlier works,\cite{BerQuaSie03,BerLauQua05} however, no previous study of such ions is available. Here we present PV calculations for CHDBrI$^+$, depicted in \Cref{fig:ChDBrI}, showing surprisingly large PV effects in several of the (chiral) vibrational modes.
CHDBrI$^+$ was proposed as a candidate for precision spectroscopy in Landau et al.\cite{landau}
In stark contrast to exotic designer molecules that contain heavy atoms,\cite{BasSch03,BasSch04,FigSauSch10,Wormit2014,marit22} the CH$_2$BrI molecule family plays an important role in atmospheric chemistry and  is naturally produced in plankton.\cite{Lim2017} 
The PV effect of CHDBrI$^+$ is by far the largest ever computed amongst such molecules \cite{Berger2000,Berger2001} and therefore is of great significance toward gauging the importance of the weak interaction in chemistry in addition to being a promising candidate for future PV measurement using trapped ion laser spectroscopy.


Within the Born--Oppenheimer approximation, the PV energy shift $E^\text{PV}=\bra{\Psi^\text{el}}\hat{H}^\text{PV}\ket{\Psi^\text{el}}=0$ for an achiral molecule such as CH$_2$BrI in its equilibrium geometry. Here $\hat{H}_\text{PV}$ is the nuclear spin-independent PV operator (see Section S1 -- Computational Details in ESI$^\dag$) and $\Psi^\text{el}$ the electronic wavefunction. 
For the isotopically chiral molecules CHDBrI and CHDBrI$^+$ considered here the electronic PV energy difference $\Delta E^\text{PV} = E^\text{PV}_\text{R}-E^\text{PV}_\text{S}$ is predicted to be exceedingly small, i.e. no more than a few mHz (and only when finite nuclear models are used). 
The situation can change dramatically, however, if we take vibrational effects into account, i.e. if we consider the expectation value $\bra{\Psi^\text{el}\Psi^\text{vib}}\hat{H}^\text{PV}\ket{\Psi^\text{el}\Psi^\text{vib}}$, which may be non-zero even for an achiral molecule for modes that break mirror image symmetry and thus induce chirality.

In order to demonstrate the importance of such vibrationally induced PV effects we studied the molecules CHDBrI and CHDBrI$^+$ using relativistic density functional theory. 
The latter system was proposed as a promising candidate for future high-precision laser spectroscopy experiments on trapped ions, due to a number of experimental advantages.\cite{landau} This proposal follows recent successful precision measurements on ionic systems, such as the molecular ion eEDM (electron electric dipole moment) measurement,\cite{RouCalWri22} and quantum logic spectroscopy of molecular ions.\cite{ColSchLei22,ChoColKur20}

\begin{table*}[ht!]
\begin{center}
\caption{Fundamental vibrational transition frequencies $\nu$ (in cm$^{-1}$), infrared intensities $I$ (in km/mol) and lifetimes $\tau$ from a harmonic analysis and parity violation shifts $\Delta\nu^\text{PV}$ (in Hz) for CHDBrI and CHDBrI$^+$.}\label{tab:Freq}
\begin{tabular}{lcccccccc}	
\hline		
Mode & CHDBrI  &  & & CHDBrI$^+$ &  & & CHDBrI	& CHDBrI$^+$ \\
& $\nu$  & $I$ & $\tau$ & $\nu$ & $I$ & $\tau$ & $\Delta\nu^\text{PV}$ & $\Delta\nu^\text{PV}$ \\
\hline
1 CBrI scis & \phantom{0}146 & \phantom{0}0.1 & $>$\phantom{.}30 min & \phantom{0}149 & \phantom{0}4.4 & 82.2 s\phantom{m} & $-$0.001 &   +0.009 \\
2 CI str    & \phantom{0}540 & \phantom{0}1.5 & 17.8 s\phantom{m} & \phantom{0}524 &           21.8 & 1.33 s\phantom{m} & $-$0.044 & $-$0.126 \\
3 CBr str   & \phantom{0}645 &           37.7 & \phantom{.}510 ms & \phantom{0}620 & \phantom{0}6.5 & 3.21 s\phantom{m} &   +0.242 & $-$0.037 \\
4 CHD rock  & \phantom{0}679 & \phantom{0}2.4 & 7.24 s\phantom{m} & \phantom{0}710 & \phantom{0}2.3 & 6.94 s\phantom{m} & $-$0.051 & $-$0.041 \\
5 CD wag    & \phantom{0}845 &           35.7 & \phantom{.}314 ms & \phantom{0}797 & \phantom{0}2.1 & 6.15 s\phantom{m} & $-$0.061 &   +1.335 \\
6 CH wag    &           1156 &           47.3 & \phantom{.}126 ms &           1104 & \phantom{0}0.8 & 8.79 s\phantom{m} &   +0.053 & $-$1.779 \\
7 CHD scis  &           1284 & \phantom{0}1.0 & 4.82 s\phantom{m} &           1275 & \phantom{0}1.8 & 2.75 s\phantom{m} &   +0.043 & $-$0.590 \\
8 CD str    &           2346 & \phantom{0}1.0 & 1.41 s\phantom{m} &           2354 & \phantom{0}7.6 & \phantom{.}191 ms &   +0.018 &   +0.937 \\
9 CH str    &           3198 & \phantom{0}2.1 & \phantom{.}368 ms &           3214 &           20.7 & 37.3 ms           & $-$0.055 & $-$1.292 \\
\hline
\end{tabular}
\end{center}
\end{table*}

\begin{figure}
\begin{center}
\includegraphics[scale=0.9]{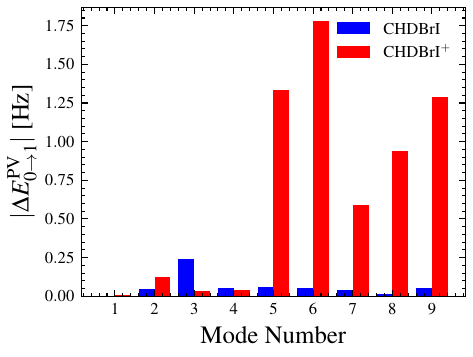}
\caption{Overview of the absolute values of PV frequency shifts $\Delta \nu^\text{PV}_{0\rightarrow 1}$ between the two enantiomers of CHDClBr and of CHDClBr$^+$ for all 9 fundamental vibrational transitions.}
\label{fig:pv_modes}
\end{center}
\end{figure}

\begin{figure}
\begin{center}
\includegraphics[scale=0.9]{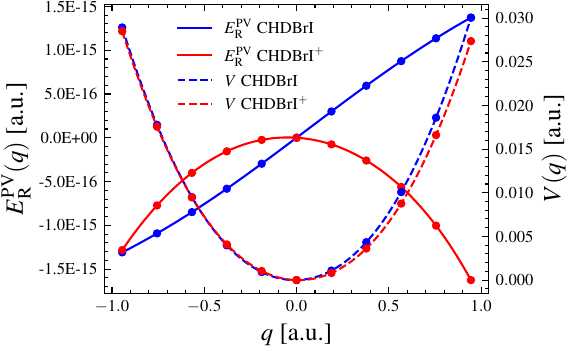}
\caption{$E^\text{PV}(q)$ and $V(q)$ curves along mode 6 of the (R)-enantiomer of CHDClBr and CHDClBr$^+$.}
\label{fig:pv_curve}
\end{center}
\end{figure}

Calculated vibrational frequencies, together with the parity violating energy contributions for both CHDBrI and CHDBrI$^+$ are shown in \Cref{tab:Freq}. 
The PV energy shifts of the fundamental transitions are summarized in \Cref{fig:pv_modes}. The vibrational PV frequency shifts between the two enantiomers are considerably enhanced in the positively charged ion compared to the neutral molecule. In particular, mode number 6 in the ion has the largest PV frequency shift compared to the other modes. In order to shine some light on the more than 30-fold enhancement in the cation, we plot the potential energy and the PV energy curves along the normal coordinate for mode 6 in \Cref{fig:pv_curve}. While the two potential energy curves for CHDBrI and CHDBrI$^+$ are almost identical and very harmonic in nature, the PV curves along the normal mode are markedly different. For the neutral case, we see an almost linear behavior with a negative and positive curvature to the left and right around $q=0$ (the equilibrium geometry) where the PV curve has a turning point, implying that $\partial^2 E^\text{PV}(q)/\partial q^2\big|_{q=0}\approx0$. Moreover, the anharmonicity is very small for mode 6. From vibrational perturbation theory (VPT) analysis (see Section S1 -- Computational Details in ESI$^\dag$) this results in a tiny PV frequency shift (with curvature and anharmonic VPT contributions of 49 and --20 mHz, respectively). In contrast, for the positively charged molecule, we see a strong curvature in the PV curve, which yields a very large contribution from the $\partial^2 E^\text{PV}(q)/\partial q^2\big|_{q\approx 0}$ term of --1.84~Hz in Equation (S2) in ESI,$^\dag$ compared to only 2~mHz for the anharmonicity term. Hence, the PV frequency shift in the cation is two orders of magnitude higher compared to the neutral molecule(!). 
Notably, the radiative lifetime of mode number 6 is found to be on the scale of seconds, such that its natural line width is narrower than the magnitude of the PV shift, and can be taken advantage of in a trapped molecular ion precision spectroscopy experiment.

In both molecules, the major atomic PV contributions come from the Br and I atoms as one expects (see Figure~S1 in ESI$^\dag$). The molecular orbital (MO) decomposition of the PV contributions reveals near-perfect cancellation between the HOMO and the rest of the MOs in the neutral molecules. The large vibrational PV effect of the cation originates from halving the PV energy contribution of the HOMO orbital (composed mainly of Br and I p-orbitals) and thus reducing the cancellation effects which are observed in the neutral system where the HOMO orbital is doubly occupied (see Section~S4 in ESI$^\dag$ for details and analysis).

\begin{figure}
\begin{center}
\includegraphics[scale=.9]{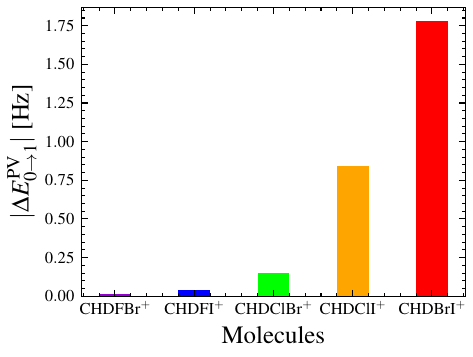}
\caption{Vibrational frequency shifts $|\Delta E^\text{PV}_{0\rightarrow 1}|$ in the different CHDXY$^+$ molecular ions (X,Y = halogens). For all systems, the results for the CH wagging mode are shown.}
\label{fig:compare_halogen}
\end{center}
\end{figure}

Comparison between the PV frequency shifts for different halogen (X, Y) combinations in CHDXY$^+$ is shown in \Cref{fig:compare_halogen}. The results are presented for the corresponding CH wagging mode with the highest PV frequency shifts for each ion. Only the system containing both bromine and iodine yields $\Delta \nu^\text{PV}_{0\rightarrow 1}$ of over 1 Hz. 
Interestingly, in all the molecular ions considered here, the modes with the largest PV frequency shifts are bending modes lying conveniently in the CO$_2$ and Quantum Cascade laser frequency range.\cite{Bernard1997,Argence2015}

The sensitivity that can be achieved in a shot-noise limited precision measurement of PV scales as $(C\tau \sqrt{N})^{-1}$ where $C$ is the fringe contrast, $\tau$ is the coherence time and $N$ is the total number of molecules detected.\cite{Shagam2023} Molecular ions excel in long coherence times due to the ease of their trapping, making the CHDBrI$^+$ candidate appealing in addition to the large PV shift expected. Since such molecules have a neutral precursor, the molecular ion may be susceptible to predissociation in a vibrational spectroscopy experiment due to a low dissociation threshold. However, the two hydrogen isotopes in CHDBrI$^+$ give rise to a high  dissociation threshold with a value of 1.29 eV as opposed to other halogenated methanes along with the high transition state energy between enantiomers of 1.3~eV, indicating that long lifetimes should be accessible for this candidate,\cite{landau} while tunneling effects should be negligible. Moreover, the proposed state preparation and detection methods for CHDBrI$^+$ through near-threshold, state-selective photo-ionization suggest that its preparation in an internally cold state is possible, improving the contrast in a future measurement.\cite{landau} Finally, CH$_2$BrI is a commercially available product, indicating that a species-wise pure racemic sample of CHDBrI could be enriched and large values of $N$ should be expected in a single shot of the experiment.

Assuming a reasonable fringe contrast of $C=0.1$ and a modest coherence time of $\tau=100$~ms for the hydrogen wagging mode, reaching precision below 1~Hz requires detecting at least 10K molecular ions. For an experimental repetition rate of 5~Hz with a conservative 10 ions detected in a cycle at the shot-noise limit, the total averaging time of a few minutes (200~sec) is needed to reach the required precision. Naturally, technical noise as well as experimentally determined $C$, $\tau$, and the experimental number of ions in a cycle, in addition to a careful study of sources of systematic uncertainty, are needed to accurately determine the projected precision. However, this promising estimate and all the other properties discussed here make CHDBrI$^+$ an excellent candidate to observe PV in molecules for the first time. With experimental investigation of the decoherence mechanisms the precision can be substantially improved by pushing the coherence time near the seconds-long limit.


In conclusion, we demonstrated that the cation CHDBrI$^+$ exhibits a vibrational CH wagging mode with high sensitivity to PV effects and a long natural lifetime, which makes it a promising potential candidate molecule for future experimental work.
The vibrational transition frequency falls into the currently available laser frequency range and the size of the expected PV frequency shift is projected to be in the region of 1 Hz. 
The large PV frequency difference is due to the fact that the PV curve shows a rather strong curvature along the normal coordinate of the CH wagging mode in contrast to the neutral species. The introduction of a positive charge reduces cancellation effects in the valence orbitals, leading to two orders of magnitude larger PV effects than in the neutral species.

\section*{Acknowledgements}

The authors thank the Center for Information Technology of the University of Groningen for their support and for providing access to the Peregrine high-performance computing cluster. Eduardus wishes to acknowledge Indonesia Endowment Fund for Education/\textit{Lembaga Pengelola Dana Pendidikan} \text{(LPDP)} for research funding. L.F.P acknowledges the support from the Slovak Research and Development Agency (APVV-20-0098, APVV-20-0127).

\section*{Conflicts of interest}
There are no conflicts to declare.





\scriptsize{

\providecommand*{\mcitethebibliography}{\thebibliography}
\csname @ifundefined\endcsname{endmcitethebibliography}
{\let\endmcitethebibliography\endthebibliography}{}

 } 


\begin{mcitethebibliography}{40}
\providecommand*{\natexlab}[1]{#1}
\providecommand*{\mciteSetBstSublistMode}[1]{}
\providecommand*{\mciteSetBstMaxWidthForm}[2]{}
\providecommand*{\mciteBstWouldAddEndPuncttrue}
  {\def\EndOfBibitem{\unskip.}}
\providecommand*{\mciteBstWouldAddEndPunctfalse}
  {\let\EndOfBibitem\relax}
\providecommand*{\mciteSetBstMidEndSepPunct}[3]{}
\providecommand*{\mciteSetBstSublistLabelBeginEnd}[3]{}
\providecommand*{\EndOfBibitem}{}
\mciteSetBstSublistMode{f}
\mciteSetBstMaxWidthForm{subitem}
{(\emph{\alph{mcitesubitemcount}})}
\mciteSetBstSublistLabelBeginEnd{\mcitemaxwidthsubitemform\space}
{\relax}{\relax}

\bibitem[Landau \emph{et~al.}(2023)Landau, Eduardus, Behar, Wallach,
  Pa\v{s}teka, Faraji, Borschevsky, and Shagam]{landau}
A.~Landau, E.~Eduardus, D.~Behar, E.~R. Wallach, L.~F. Pa\v{s}teka, S.~Faraji,
  A.~Borschevsky and Y.~Shagam, \emph{arXiv preprint}, 2023,  2306.09788\relax
\mciteBstWouldAddEndPuncttrue
\mciteSetBstMidEndSepPunct{\mcitedefaultmidpunct}
{\mcitedefaultendpunct}{\mcitedefaultseppunct}\relax
\EndOfBibitem
\bibitem[Pasteur(1874)]{Pasteur}
L.~Pasteur, \emph{Comptes Rendus de l'Académie des Sciences}, 1874,
  \textbf{78}, 1515--1518\relax
\mciteBstWouldAddEndPuncttrue
\mciteSetBstMidEndSepPunct{\mcitedefaultmidpunct}
{\mcitedefaultendpunct}{\mcitedefaultseppunct}\relax
\EndOfBibitem
\bibitem[Ginges and Flambaum(2004)]{Ginges2004}
J.~Ginges and V.~Flambaum, \emph{Physics Reports}, 2004, \textbf{397}, 63 --
  154\relax
\mciteBstWouldAddEndPuncttrue
\mciteSetBstMidEndSepPunct{\mcitedefaultmidpunct}
{\mcitedefaultendpunct}{\mcitedefaultseppunct}\relax
\EndOfBibitem
\bibitem[Jungmann(2014)]{Jungmann2014}
K.~P. Jungmann, \emph{Hyperfine Interactions}, 2014, \textbf{228}, 21--29\relax
\mciteBstWouldAddEndPuncttrue
\mciteSetBstMidEndSepPunct{\mcitedefaultmidpunct}
{\mcitedefaultendpunct}{\mcitedefaultseppunct}\relax
\EndOfBibitem
\bibitem[Mason(1988)]{Mason1988}
S.~Mason, \emph{Chemical Society Reviews}, 1988, \textbf{17}, 347--359\relax
\mciteBstWouldAddEndPuncttrue
\mciteSetBstMidEndSepPunct{\mcitedefaultmidpunct}
{\mcitedefaultendpunct}{\mcitedefaultseppunct}\relax
\EndOfBibitem
\bibitem[Bonner(1995)]{Bonner1995}
W.~A. Bonner, \emph{Origins of Life and Evolution of the Biosphere}, 1995,
  \textbf{25}, 175--190\relax
\mciteBstWouldAddEndPuncttrue
\mciteSetBstMidEndSepPunct{\mcitedefaultmidpunct}
{\mcitedefaultendpunct}{\mcitedefaultseppunct}\relax
\EndOfBibitem
\bibitem[Blackmond(2020)]{Blackmond2020}
D.~G. Blackmond, \emph{Chemical Reviews}, 2020, \textbf{120}, 4831--4847\relax
\mciteBstWouldAddEndPuncttrue
\mciteSetBstMidEndSepPunct{\mcitedefaultmidpunct}
{\mcitedefaultendpunct}{\mcitedefaultseppunct}\relax
\EndOfBibitem
\bibitem[Wu \emph{et~al.}(1957)Wu, Ambler, Hayward, Hoppes, and Hudson]{Wu1}
C.~S. Wu, E.~Ambler, R.~W. Hayward, D.~D. Hoppes and R.~P. Hudson,
  \emph{Physical Review}, 1957, \textbf{105}, 1413--1415\relax
\mciteBstWouldAddEndPuncttrue
\mciteSetBstMidEndSepPunct{\mcitedefaultmidpunct}
{\mcitedefaultendpunct}{\mcitedefaultseppunct}\relax
\EndOfBibitem
\bibitem[Letokhov(1975)]{Letokhov75}
V.~Letokhov, \emph{Physics Letters A}, 1975, \textbf{53}, 275--276\relax
\mciteBstWouldAddEndPuncttrue
\mciteSetBstMidEndSepPunct{\mcitedefaultmidpunct}
{\mcitedefaultendpunct}{\mcitedefaultseppunct}\relax
\EndOfBibitem
\bibitem[Laerdahl \emph{et~al.}(2000)Laerdahl, Wesendrup, and
  Schwerdtfeger]{Laerdahl2000a}
J.~K. Laerdahl, R.~Wesendrup and P.~Schwerdtfeger, \emph{ChemPhysChem}, 2000,
  \textbf{1}, 60--62\relax
\mciteBstWouldAddEndPuncttrue
\mciteSetBstMidEndSepPunct{\mcitedefaultmidpunct}
{\mcitedefaultendpunct}{\mcitedefaultseppunct}\relax
\EndOfBibitem
\bibitem[Wesendrup \emph{et~al.}(2003)Wesendrup, Laerdahl, Compton, and
  Schwerdtfeger]{Wesendrup2003}
R.~Wesendrup, J.~K. Laerdahl, R.~N. Compton and P.~Schwerdtfeger, \emph{The
  Journal of Physical Chemistry A}, 2003, \textbf{107}, 6668--6673\relax
\mciteBstWouldAddEndPuncttrue
\mciteSetBstMidEndSepPunct{\mcitedefaultmidpunct}
{\mcitedefaultendpunct}{\mcitedefaultseppunct}\relax
\EndOfBibitem
\bibitem[Mart{\'\i}nez \emph{et~al.}(2022)Mart{\'\i}nez, Cuccia, Viedma, and
  Cintas]{Cintas2022}
R.~F. Mart{\'\i}nez, L.~A. Cuccia, C.~Viedma and P.~Cintas, \emph{Origins of
  Life and Evolution of Biospheres}, 2022, \textbf{52}, 21--56\relax
\mciteBstWouldAddEndPuncttrue
\mciteSetBstMidEndSepPunct{\mcitedefaultmidpunct}
{\mcitedefaultendpunct}{\mcitedefaultseppunct}\relax
\EndOfBibitem
\bibitem[Sallembien \emph{et~al.}(2022)Sallembien, Bouteiller, Crassous, and
  Raynal]{Crassous2022}
Q.~Sallembien, L.~Bouteiller, J.~Crassous and M.~Raynal, \emph{Chemical Society
  Reviews}, 2022, \textbf{51}, 3436--3476\relax
\mciteBstWouldAddEndPuncttrue
\mciteSetBstMidEndSepPunct{\mcitedefaultmidpunct}
{\mcitedefaultendpunct}{\mcitedefaultseppunct}\relax
\EndOfBibitem
\bibitem[Crassous \emph{et~al.}(2005)Crassous, Chardonnet, Saue, and
  Schwerdtfeger]{CraChaSau05}
J.~Crassous, C.~Chardonnet, T.~Saue and P.~Schwerdtfeger, \emph{Organic and
  Biomolecular Chemistry}, 2005, \textbf{3}, 2218--2224\relax
\mciteBstWouldAddEndPuncttrue
\mciteSetBstMidEndSepPunct{\mcitedefaultmidpunct}
{\mcitedefaultendpunct}{\mcitedefaultseppunct}\relax
\EndOfBibitem
\bibitem[Darqui\'e \emph{et~al.}(2010)Darqui\'e, Stoeffler, Zrig, Crassous,
  Soulard, Asselin, Huet, Guy, Bast, Saue, Schwerdtfeger, Shelkovnikov, Daussy,
  Amy-Klein, and Chardonnet]{DarStoZri10}
B.~Darqui\'e, C.~Stoeffler, S.~Zrig, J.~Crassous, P.~Soulard, P.~Asselin, T.~R.
  Huet, L.~Guy, R.~Bast, T.~Saue, P.~Schwerdtfeger, A.~Shelkovnikov, C.~Daussy,
  A.~Amy-Klein and C.~Chardonnet, \emph{Chirality}, 2010, \textbf{22},
  870--884\relax
\mciteBstWouldAddEndPuncttrue
\mciteSetBstMidEndSepPunct{\mcitedefaultmidpunct}
{\mcitedefaultendpunct}{\mcitedefaultseppunct}\relax
\EndOfBibitem
\bibitem[Quack \emph{et~al.}(2022)Quack, Seyfang, and Wichmann]{Quack2022}
M.~Quack, G.~Seyfang and G.~Wichmann, \emph{Chemical Science}, 2022,
  \textbf{13}, 10598--10643\relax
\mciteBstWouldAddEndPuncttrue
\mciteSetBstMidEndSepPunct{\mcitedefaultmidpunct}
{\mcitedefaultendpunct}{\mcitedefaultseppunct}\relax
\EndOfBibitem
\bibitem[Daussy \emph{et~al.}(1999)Daussy, Marrel, Amy-Klein, Nguyen, Bord\'e,
  and Chardonnet]{DauMarAmy99}
C.~Daussy, T.~Marrel, A.~Amy-Klein, C.~T. Nguyen, C.~J. Bord\'e and
  C.~Chardonnet, \emph{Physical Review Letters}, 1999, \textbf{83},
  1554--1557\relax
\mciteBstWouldAddEndPuncttrue
\mciteSetBstMidEndSepPunct{\mcitedefaultmidpunct}
{\mcitedefaultendpunct}{\mcitedefaultseppunct}\relax
\EndOfBibitem
\bibitem[Crassous \emph{et~al.}(2003)Crassous, Monier, Dutasta, Ziskind,
  Daussy, Grain, and Chardonnet]{chardonnet:newexp}
J.~Crassous, J.-P. Monier, F.~Dutasta, M.~Ziskind, C.~Daussy, C.~Grain and
  C.~Chardonnet, \emph{ChemPhysChem}, 2003, \textbf{4}, 541\relax
\mciteBstWouldAddEndPuncttrue
\mciteSetBstMidEndSepPunct{\mcitedefaultmidpunct}
{\mcitedefaultendpunct}{\mcitedefaultseppunct}\relax
\EndOfBibitem
\bibitem[Szab{\'o}-Nagy and Keszthelyi(1999)]{szabo1999}
A.~Szab{\'o}-Nagy and L.~Keszthelyi, \emph{Proceedings of the National Academy
  of Sciences}, 1999, \textbf{96}, 4252--4255\relax
\mciteBstWouldAddEndPuncttrue
\mciteSetBstMidEndSepPunct{\mcitedefaultmidpunct}
{\mcitedefaultendpunct}{\mcitedefaultseppunct}\relax
\EndOfBibitem
\bibitem[Wang \emph{et~al.}(2000)Wang, Yi, Ni, Zhao, Jin, and Tang]{Wang2000}
W.~Wang, F.~Yi, Y.~Ni, Z.~Zhao, X.~Jin and Y.~Tang, \emph{Journal of Biological
  Physics}, 2000, \textbf{26}, 51--65\relax
\mciteBstWouldAddEndPuncttrue
\mciteSetBstMidEndSepPunct{\mcitedefaultmidpunct}
{\mcitedefaultendpunct}{\mcitedefaultseppunct}\relax
\EndOfBibitem
\bibitem[Wang \emph{et~al.}(2003)Wang, Min, Zhu, and Yi]{Wang2003}
W.~Wang, W.~Min, C.~Zhu and F.~Yi, \emph{Physical Chemistry Chemical Physics},
  2003, \textbf{5}, 4000--4003\relax
\mciteBstWouldAddEndPuncttrue
\mciteSetBstMidEndSepPunct{\mcitedefaultmidpunct}
{\mcitedefaultendpunct}{\mcitedefaultseppunct}\relax
\EndOfBibitem
\bibitem[Bast and Schwerdtfeger(2003)]{BasSch03}
R.~Bast and P.~Schwerdtfeger, \emph{Physical Review Letters}, 2003,
  \textbf{91}, 023001\relax
\mciteBstWouldAddEndPuncttrue
\mciteSetBstMidEndSepPunct{\mcitedefaultmidpunct}
{\mcitedefaultendpunct}{\mcitedefaultseppunct}\relax
\EndOfBibitem
\bibitem[Schwerdtfeger and Bast(2004)]{BasSch04}
P.~Schwerdtfeger and R.~Bast, \emph{Journal of the American Chemical Society},
  2004, \textbf{126}, 1652--1653\relax
\mciteBstWouldAddEndPuncttrue
\mciteSetBstMidEndSepPunct{\mcitedefaultmidpunct}
{\mcitedefaultendpunct}{\mcitedefaultseppunct}\relax
\EndOfBibitem
\bibitem[Figgen \emph{et~al.}(2010)Figgen, Saue, and
  Schwerdtfeger]{FigSauSch10}
D.~Figgen, T.~Saue and P.~Schwerdtfeger, \emph{The Journal of Chemical
  Physics}, 2010, \textbf{132}, 234310\relax
\mciteBstWouldAddEndPuncttrue
\mciteSetBstMidEndSepPunct{\mcitedefaultmidpunct}
{\mcitedefaultendpunct}{\mcitedefaultseppunct}\relax
\EndOfBibitem
\bibitem[Wormit \emph{et~al.}(2014)Wormit, Olejniczak, Deppenmeier,
  Borschevsky, Saue, and Schwerdtfeger]{Wormit2014}
M.~Wormit, M.~Olejniczak, A.-L. Deppenmeier, A.~Borschevsky, T.~Saue and
  P.~Schwerdtfeger, \emph{Phys. Chem. Chem. Phys.}, 2014, \textbf{16},
  17043--17051\relax
\mciteBstWouldAddEndPuncttrue
\mciteSetBstMidEndSepPunct{\mcitedefaultmidpunct}
{\mcitedefaultendpunct}{\mcitedefaultseppunct}\relax
\EndOfBibitem
\bibitem[Fiechter \emph{et~al.}(2022)Fiechter, Haase, Saleh, Soulard, Tremblay,
  Havenith, Timmermans, Schwerdtfeger, Crassous, Darquié, Pašteka, and
  Borschevsky]{marit22}
M.~R. Fiechter, P.~A.~B. Haase, N.~Saleh, P.~Soulard, B.~Tremblay, R.~W.~A.
  Havenith, R.~G.~E. Timmermans, P.~Schwerdtfeger, J.~Crassous, B.~Darquié,
  L.~F. Pašteka and A.~Borschevsky, \emph{The Journal of Physical Chemistry
  Letters}, 2022, \textbf{13}, 10011--10017\relax
\mciteBstWouldAddEndPuncttrue
\mciteSetBstMidEndSepPunct{\mcitedefaultmidpunct}
{\mcitedefaultendpunct}{\mcitedefaultseppunct}\relax
\EndOfBibitem
\bibitem[Bouchiat and Bouchiat(1974)]{Bouchiat:1974:physlettb}
M.~A. Bouchiat and C.~C. Bouchiat, \emph{Phys. Lett. B}, 1974, \textbf{48},
  111--114\relax
\mciteBstWouldAddEndPuncttrue
\mciteSetBstMidEndSepPunct{\mcitedefaultmidpunct}
{\mcitedefaultendpunct}{\mcitedefaultseppunct}\relax
\EndOfBibitem
\bibitem[Laerdahl and Schwerdtfeger(1999)]{LaeSch99}
J.~K. Laerdahl and P.~Schwerdtfeger, \emph{Phys. Rev. A}, 1999, \textbf{60},
  4439--4453\relax
\mciteBstWouldAddEndPuncttrue
\mciteSetBstMidEndSepPunct{\mcitedefaultmidpunct}
{\mcitedefaultendpunct}{\mcitedefaultseppunct}\relax
\EndOfBibitem
\bibitem[Buda \emph{et~al.}(1992)Buda, der Heyde, and Mislow]{Mislow1992}
A.~B. Buda, T.~A. der Heyde and K.~Mislow, \emph{Angewandte Chemie
  International Edition}, 1992, \textbf{31}, 989--1007\relax
\mciteBstWouldAddEndPuncttrue
\mciteSetBstMidEndSepPunct{\mcitedefaultmidpunct}
{\mcitedefaultendpunct}{\mcitedefaultseppunct}\relax
\EndOfBibitem
\bibitem[Berger \emph{et~al.}(2003)Berger, Quack, Sieben, and
  Willeke]{BerQuaSie03}
R.~Berger, M.~Quack, A.~Sieben and M.~Willeke, \emph{Helvetica Chimica Acta},
  2003, \textbf{86}, 4048--4060\relax
\mciteBstWouldAddEndPuncttrue
\mciteSetBstMidEndSepPunct{\mcitedefaultmidpunct}
{\mcitedefaultendpunct}{\mcitedefaultseppunct}\relax
\EndOfBibitem
\bibitem[Berger \emph{et~al.}(2005)Berger, Laubender, Quack, Sieben, Stohner,
  and Willeke]{BerLauQua05}
R.~Berger, G.~Laubender, M.~Quack, A.~Sieben, J.~Stohner and M.~Willeke,
  \emph{Angewandte Chemie International Edition}, 2005, \textbf{44},
  3623--3626\relax
\mciteBstWouldAddEndPuncttrue
\mciteSetBstMidEndSepPunct{\mcitedefaultmidpunct}
{\mcitedefaultendpunct}{\mcitedefaultseppunct}\relax
\EndOfBibitem
\bibitem[Lim \emph{et~al.}(2017)Lim, Phang, Rahman, Sturges, and
  Malin]{Lim2017}
Y.~K. Lim, S.~M. Phang, N.~A. Rahman, W.~T. Sturges and G.~Malin,
  \emph{International Journal of Environmental Science and Technology}, 2017,
  \textbf{14}, 1355--1370\relax
\mciteBstWouldAddEndPuncttrue
\mciteSetBstMidEndSepPunct{\mcitedefaultmidpunct}
{\mcitedefaultendpunct}{\mcitedefaultseppunct}\relax
\EndOfBibitem
\bibitem[Berger and Quack(2000)]{Berger2000}
R.~Berger and M.~Quack, \emph{ChemPhysChem}, 2000, \textbf{1}, 57--60\relax
\mciteBstWouldAddEndPuncttrue
\mciteSetBstMidEndSepPunct{\mcitedefaultmidpunct}
{\mcitedefaultendpunct}{\mcitedefaultseppunct}\relax
\EndOfBibitem
\bibitem[Berger \emph{et~al.}(2001)Berger, Quack, and Stohner]{Berger2001}
R.~Berger, M.~Quack and J.~Stohner, \emph{Angewandte Chemie International
  Edition}, 2001, \textbf{40}, 1661\relax
\mciteBstWouldAddEndPuncttrue
\mciteSetBstMidEndSepPunct{\mcitedefaultmidpunct}
{\mcitedefaultendpunct}{\mcitedefaultseppunct}\relax
\EndOfBibitem
\bibitem[Roussy \emph{et~al.}(2022)Roussy, Caldwell, Wright, Cairncross,
  Shagam, Ng, Schlossberger, Park, Wang, Ye, and Cornell]{RouCalWri22}
T.~S. Roussy, L.~Caldwell, T.~Wright, W.~B. Cairncross, Y.~Shagam, K.~B. Ng,
  N.~Schlossberger, S.~Y. Park, A.~Wang, J.~Ye and E.~A. Cornell, \emph{arXiv
  preprint}, 2022,  2212.11841\relax
\mciteBstWouldAddEndPuncttrue
\mciteSetBstMidEndSepPunct{\mcitedefaultmidpunct}
{\mcitedefaultendpunct}{\mcitedefaultseppunct}\relax
\EndOfBibitem
\bibitem[Collopy \emph{et~al.}(2022)Collopy, Schmidt, Leibfried, Leibrandt, and
  Chou]{ColSchLei22}
A.~L. Collopy, J.~Schmidt, D.~Leibfried, D.~R. Leibrandt and C.-W. Chou,
  \emph{arXiv preprint}, 2022,  2207.10215\relax
\mciteBstWouldAddEndPuncttrue
\mciteSetBstMidEndSepPunct{\mcitedefaultmidpunct}
{\mcitedefaultendpunct}{\mcitedefaultseppunct}\relax
\EndOfBibitem
\bibitem[Chou \emph{et~al.}(2020)Chou, Collopy, Kurz, Lin, Harding, Plessow,
  Fortier, Diddams, Leibfried, and Leibrandt]{ChoColKur20}
C.-W. Chou, A.~L. Collopy, C.~Kurz, Y.~Lin, M.~E. Harding, P.~N. Plessow,
  T.~Fortier, S.~Diddams, D.~Leibfried and D.~R. Leibrandt, \emph{Science},
  2020, \textbf{367}, 1458--1461\relax
\mciteBstWouldAddEndPuncttrue
\mciteSetBstMidEndSepPunct{\mcitedefaultmidpunct}
{\mcitedefaultendpunct}{\mcitedefaultseppunct}\relax
\EndOfBibitem
\bibitem[Bernard \emph{et~al.}(1997)Bernard, Daussy, Nogues, Constantin,
  Durand, Amy-Klein, Lerberghe, and Chardonnet]{Bernard1997}
V.~Bernard, C.~Daussy, G.~Nogues, L.~Constantin, P.~E. Durand, A.~Amy-Klein,
  A.~V. Lerberghe and C.~Chardonnet, \emph{IEEE Journal of Quantum
  Electronics}, 1997, \textbf{33}, 1282--1287\relax
\mciteBstWouldAddEndPuncttrue
\mciteSetBstMidEndSepPunct{\mcitedefaultmidpunct}
{\mcitedefaultendpunct}{\mcitedefaultseppunct}\relax
\EndOfBibitem
\bibitem[Argence \emph{et~al.}(2015)Argence, Chanteau, Lopez, Nicolodi,
  Abgrall, Chardonnet, Daussy, Darquié, Coq, and Amy-Klein]{Argence2015}
B.~Argence, B.~Chanteau, O.~Lopez, D.~Nicolodi, M.~Abgrall, C.~Chardonnet,
  C.~Daussy, B.~Darquié, Y.~L. Coq and A.~Amy-Klein, \emph{Nature Photonics},
  2015, \textbf{9}, 456--460\relax
\mciteBstWouldAddEndPuncttrue
\mciteSetBstMidEndSepPunct{\mcitedefaultmidpunct}
{\mcitedefaultendpunct}{\mcitedefaultseppunct}\relax
\EndOfBibitem
\bibitem[Erez \emph{et~al.}(2023)Erez, Wallach, and Shagam]{Shagam2023}
I.~Erez, E.~R. Wallach and Y.~Shagam, \emph{arXiv preprint}, 2023,
  2206.03699\relax
\mciteBstWouldAddEndPuncttrue
\mciteSetBstMidEndSepPunct{\mcitedefaultmidpunct}
{\mcitedefaultendpunct}{\mcitedefaultseppunct}\relax
\EndOfBibitem
\end{mcitethebibliography}
\end{document}